# Photodetachment Spectroscopy of Quasibound States of the Negative Ion of Lanthanum

C. W. Walter[*], N. D. Gibson, N. B. Lyman[†], J. Wang[#]

*Department of Physics and Astronomy, Denison University, Granville, Ohio 43023, USA*

(Dated: September 23, 2020)

## ABSTRACT

The negative ion of lanthanum, La¯, has one of the richest bound state spectra observed for an atomic negative ion and has been proposed as a promising candidate for laser-cooling applications. In the present experiments, La¯ was investigated using tunable infrared photodetachment spectroscopy. The relative signal for neutral atom production was measured with a crossed ion-beam--laser-beam apparatus over the photon energy range 590 - 920 meV (2100 - 1350 nm) to probe the continuum region above the La neutral atom ground state. Eleven prominent peaks were observed in the La¯ photodetachment cross section due to resonant excitation of quasibound transient negative ion states in the continuum which subsequently autodetach. In addition, thresholds were observed for photodetachment from several bound states of La¯ to both ground and excited states of La. The present results provide information on the excited state structure and dynamics of La¯ that depend crucially on multielectron correlation effects.

---

[*]Electronic address: walter@denison.edu.

[†]Present address: Boeing Satellite Systems, 1950 E Imperial Hwy, El Segundo, CA 90245, USA.

[#]Present address: Department of Physics and Astronomy, University of Nebraska-Lincoln, Lincoln, NE 68588, USA.

**I. Introduction**

Among all atomic negative ions, that of lanthanum stands out as among the most intriguing. La¯ has one of the richest absorption spectra of highly unusual bound-bound electric dipole transitions observed for an atomic negative ion [1]. Furthermore, La¯ has emerged as one of the most promising negative ions for laser-cooling applications [1-5]. Given the unique aspects of the lanthanum negative ion, further investigation is clearly needed, and the present study explores the energy region above the photodetachment threshold of La¯, revealing new resonance and threshold structures.

The complexity of the lanthanum negative ion is due to the existence of several open subshells ($6p$, $5d$, and $4f$) that lie close in energy in neutral La (ground state [Xe]$5d6s^2$ $^2D_{3/2}$). Therefore, multielectron interactions and core-valence correlations are crucial for its structure and properties. The uniqueness of La¯ was recognized in the 1990's through theoretical investigations by Vosko *et al.* [6] and later by O'Malley and Beck [7] that predicted that La¯ has multiple bound states of opposite parity formed by attachment of either a $6p$ or $5d$ electron to neutral La. Furthermore, initial photoelectron spectroscopy experiments on La¯ carried out by Covington *et al.* indicated the existence of at least two bound states [8].

The understanding of La¯ has progressed substantially over the past decade through both theoretical and experimental investigations. In 2009, O'Malley and Beck performed detailed relativistic configuration interaction calculations indicating that La¯ has 7 even-parity bound states with configuration [Xe]$5d^26s^2$ and 8 odd-parity bound states with configuration [Xe]$5d6s^26p$ [9]. Their group then did further calculations of transitions between bound states of the negative ion and suggested one transition, La¯ $^3F_2^e \rightarrow {}^3D_1^o$, as a promising candidate for laser-cooling [2,3]. Subsequently, our group used photodetachment spectroscopy to observe and identify 12 electric dipole transitions between bound states of La¯, confirming the general theoretical predictions and greatly refining the state excitation energies [1]. Such a rich bound state spectrum is very unusual for atomic negative ions, which typically only have a single bound-state configuration because the extra electron is not bound by a net long-range Coulomb field. Kellerbauer and co-workers then made more detailed spectroscopic measurements for two of the bound-bound transitions [4,5,10] as well as additional theoretical calculations of the bound state structure and transition rates [5]; these studies confirmed La¯ as a promising candidate for laser cooling. Very recently, Lu *et al.* [11] used slow electron velocity map imaging to measure the binding energies of all of the bound states of La¯, including determining the electron affinity of La as 557.553(25) meV. Subsequently, Blondel made a detailed analysis of the complete set of all available experimental data on La¯ [1,4,5,10,11] to obtain a slightly revised La electron affinity of 557.546(20) meV, as well as optimized binding energies for all of the bound states of La¯ [12].

In contrast to the intense research efforts on the bound state structure of La¯, relatively little previous work has explored its photodetachment spectrum. Our group's previous study of La¯ focused on bound-bound transitions, with only limited exploration of the region above the ground state La threshold [1]. In addition, the fixed-frequency photoelectron spectroscopy experiments by Covington *et al.* [8] and Lu *et al.* [11] yield some information about relative photodetachment channel strengths. On the theory side, Pan and Beck have performed calculations of photodetachment cross sections from La¯ at two discrete photon energies: 337 meV to assess photodetachment as a loss mechanism from the $^3D_1^o$ upper-level of the potential laser-cooling transition [3]; and 2410 meV [13] to re-interpret the earlier photoelectron spectroscopy data of Covington *et al.* [8]. However, none of the previous studies have investigated the photon energy dependence of the continuum photodetachment spectrum or the quasibound resonance structure of La¯.

In the present study, photodetachment spectroscopy of La¯ with a tunable mid-infrared laser was used to investigate the region above the neutral La threshold from 585 - 920 meV. The relevant states of La¯ and La are shown in the energy level diagram of Fig. 1. At least 11 prominent peaks were observed in the La¯ photodetachment cross section in this range due to resonant excitation of quasibound transient negative ion states in the continuum. In addition, photodetachment thresholds were observed for some opening channels. The present results provide further information on the excited state structure and dynamics of the intriguing lanthanum negative ion.

**II. Experimental Method**

In the present experiments, photodetachment from La¯ was measured as a function of photon energy using a crossed ion-beam—laser-beam system that has been described in detail previously [14,15]. Negative ions were produced by a cesium sputter ion source (NEC SNICS II) [16] using a cathode packed with powdered lanthanum oxide covered by a tungsten layer [17]. The ions were accelerated to 12 keV and the $^{139}$La¯ isotope was magnetically mass selected; typical currents of La¯ were ~30 pA. In the interaction region, the ion beam was intersected perpendicularly by a pulsed laser beam. Following the interaction region, residual negative ions were electrostatically deflected into a Faraday cup, while neutral atoms produced by photodetachment continued undeflected to a multidynode particle multiplier detector. The neutral atom signal was normalized to the ion-beam current and the laser photon flux measured for each laser pulse. The spectra were obtained by repeatedly continuously scanning the laser wavelength over a range and then sorting the data into photon energy bins of selectable width, as previously described by Walter *et al.* [14].

The laser system consisted of a tunable optical parametric oscillator-amplifier (OPO-OPA) (LaserVision) pumped by a pulsed Nd:YAG laser operating at 20 Hz. Both the "signal" and "idler" output bands of the OPA were used in the present measurements, giving an operating range of 250 – 920 meV (5000 – 1350 nm). Broad scans across the full tuning range were performed with the pump laser operating broadband giving an OPA bandwidth of ~0.1 meV, while narrow scans near threshold were performed with injection seeding of the pump laser to reduce the OPA bandwidth to ~0.01 meV. The wavelength of the mid-infrared light was determined for each laser pulse using a procedure fully described in [14]; briefly, a wavemeter (High Finesse WS6–600) measured the wavelength of the OPO "signal" light, which was then used with the measured pump laser wavelength to determine the wavelength of the OPA light by conservation of energy. The laser beam diverges slightly as it leaves the OPA, so a long focal length lens (~2 m focal length) was placed in the beam path ~2 m from the interaction region to approximately collimate the beam. In the interaction region, the laser pulse had a typical energy of ~50 μJ, pulse duration of ~5 ns, and beamwidth of ~0.25 cm. To reduce room air absorption by strong $H_2O$ and $CO_2$ bands in the mid-infrared [18], a tube flushed with dry nitrogen gas was used to enclose the laser beam path from the OPA to the vacuum chamber entrance window.

The sputter ion source populates many different bound states of La¯ due to its high temperature. Most of the excited states are expected to have lifetimes of tens of μs to ms [2], therefore excited ions should survive the ~25 μs flight time from source to interaction region. Thus, the target ions are in a range of initial states including the ground and a variety of excited states.

# III. Results and Discussion

The relative photodetachment cross section from La¯ measured over the photon energy range 290 – 920 meV is shown in Fig. 2; the data for photon energies below 585 meV are from Walter *et al.* [1] and the data above 585 meV were acquired in the present study. This spectrum shows slowly varying components due to non-resonant photodetachment and a number of prominent peaks due to resonant photodetachment. The non-resonant structure is due to the opening of new photodetachment channels to the ground and excited states of neutral La together with variations in the continuum photodetachment cross section as a function of photon energy. The resonance peaks are due to photoexcitation of either bound states of the negative ion or quasibound temporary states in the continuum. The threshold and resonance structures are fully discussed in the following sections.

## III.A. Threshold Structures

For a limited range above an opening threshold, the photodetachment cross section from a negative ion is characterized by the Wigner threshold law [19]:

$$\sigma = \sigma_0 + a(E - E_t)^{\ell + \frac{1}{2}} \qquad (1)$$

where $E$ is the photon energy, $E_t$ is the threshold energy, $\ell$ is the orbital angular momentum of the departing electron, and $a$ is a scaling constant. The background cross section due to photodetachment to lower energy thresholds is represented by $\sigma_0$, which may be energy dependent. In the present experiments, *s*, *p,* and *d* electrons are detached from different initial states of the La¯ ion. The angular momentum selection rule $\Delta \ell = \pm 1$ restricts the possible final angular momentum of the departing electron. In each case, the Wigner law indicates that the dominant partial wave near threshold will be that with the lowest allowed angular momentum. As the energy above threshold increases, the photodetachment cross section progressively deviates from the Wigner law due to long-range interactions between the departing electron and the remaining neutral atom, such as polarization forces [20], and the threshold function is no longer applicable.

Several photodetachment thresholds are distinguishable in the measured spectrum shown in Fig. 2. As previously reported by our group [1], the lowest energy threshold observed over this energy range is a sharp *s*-wave threshold due to detachment from the La¯ ($5d6s^26p$ $^1D_2^o$) excited state to the neutral atom ground state La ($5d6s^2$ $^2D_{3/2}$) at 335.6(8) meV, which corresponds to the binding energy of the $^1D_2^o$ state. Note that the threshold for detachment from $^1D_2^o$ to the fine structure excited level of the La ground state $^2D_{5/2}$, which would occur at an energy of 466.2 meV, is not distinctive in the spectrum of Fig. 2; this finding is consistent with the calculations of Pan and Beck, which showed a much smaller cross section for photodetachment from $^1D_2^o$ to $^2D_{5/2}$ than to the ground state $^2D_{3/2}$ [13]. It is worthwhile in the present paper to compare our previously measured binding energy of the La¯ $^1D_2^o$ state of 335.6(8) meV [1] to subsequent theoretical [5] and experimental [11] values. The theoretical calculations from Cerchiari *et al.* [5] yielded a binding energy of La¯ $^1D_2^o$ of 342(14) meV [21], which is consistent with our measured value. The photoelectron spectroscopy experiment of Lu *et al.* [11] yielded a binding energy for the La¯ $^1D_2^o$ state of 335(7) meV, which is in excellent agreement with our photodetachment threshold spectroscopy value, albeit with a larger uncertainty.

There are also several *p*-wave thresholds visible in the spectrum of Fig. 2 due to detachment from more strongly-bound states of La¯. Gradual rises in the cross section are observed over the range 560 - 720 meV above the thresholds for detachment from the ground and first fine-structure excited states of La¯ ($5d^26s^2$ $^3F_{2,3}^e$) to the ground and first fine-structure excited states of neutral La ($5d6s^2$ $^2D_{3/2,5/2}$). In this case, a *d* electron is detached, so the *p*-wave dominates the cross section near threshold giving a slow increase in the cross section of $(E - E_t)^{3/2}$ by the Wigner Law.

Although the weakness of these thresholds in comparison to the background from detachment of less strongly-bound ions precludes meaningful fitting of the measured cross section, the observed structures are clearly consistent with the expected energies of these thresholds based on the electron affinity of La [11,12] and the fine-structure splittings of La¯ [1,11,12] and La [22].

At higher energies above 800 meV, several more thresholds are observed in the spectrum due to photodetachment from the negative ion ground state manifold La¯ ($5d^26s^2$ $^3F_J^e$) to the neutral atom first excited electronic state manifold La ($5d^26s$ $^4F_J$). An expanded view of one clear threshold in this manifold, La¯ ($5d^26s^2$ $^3F_4^e$) ––> La ($5d^26s$ $^4F_{7/2}$), is shown in Fig. 3. In this case, an $s$ electron is detached, resulting in a $p$-wave detachment cross section near threshold varying as $(E - E_t)^{3/2}$ by the Wigner Law. A fit of the Wigner Law $p$-wave to the data in Fig. 3 yields a threshold energy of 819.6(19) meV, which is in agreement with the expected energy of the $^3F_4^e$ ––> $^4F_{7/2}$ threshold of 817.94(3) meV [23]. Also visible in Fig. 3 is a prominent resonance at an energy ~28 meV above the detachment threshold, which is identified as Peak 21. The thresholds at even higher energies are obscured by resonance peaks that are closer to threshold, as discussed in the next section.

### III.B. Resonances

Figure 2 shows 11 of the 12 resonance peaks in the photodetachment spectrum from La¯ observed in our previous work [1], as well as 11 newly observed peaks at higher photon energy from the present study. Two different processes cause peaks in the neutral atom spectrum: One-photon detachment via a quasibound resonance in the continuum which subsequently autodetaches, or resonance enhanced (1 + 1) photon detachment in which one photon excites the ion from a lower bound state to an upper bound state which then absorbs a second photon to detach an electron. As previously reported by our group [1], the lower energy peaks (Peaks 1 - 12) in the observed La¯ photodetachment spectrum are due to the latter (1+1) resonant detachment process through bound-bound electric dipole transitions in the negative ion. The newly observed higher energy peaks found in the present study (Peaks 13 - 23) are due to one-photon detachment via quasibound resonances in the continuum.

Traditionally, transient quasibound excited states in negative ions are classified as being either Feshbach resonances (which are formed by attachment of an electron to an excited state of the neutral atom to form a negative ion at an energy *below* the neutral atom state) or shape resonances (which are formed by trapping of the departing electron by a centrifugal barrier *above* the energy of the neutral atom state) [24,25]. Both of these types of resonances lie above the lowest detachment threshold for a particular negative ion state, thus an electron can subsequently autodetach from the photoexcited negative ion to produce a neutral atom. Feshbach resonances usually have longer lifetimes than shape resonances because they lie below their parent state, thus the spectral features associated with Feshbach resonances are generally narrower in width.

To determine the peak energies and widths of the observed resonances, Fano profiles [26] were fit to the peaks. The Fano formula gives the cross section in the vicinity of the peak as:

$$\sigma = \sigma_0 + b\frac{(q+\varepsilon)^2}{1+\varepsilon^2} \qquad (2)$$

where $\sigma_0$ is the continuum cross section (assumed constant over the narrow energy range for most of the peak fits in the present case), $q$ is the asymmetry parameter, and $b$ is a scaling constant. The factor $\varepsilon$ is given by $2(E - E_r)/\Gamma$, where $E$ is the photon energy, $E_r$ is the energy of the resonance, and $\Gamma$ is the peak width (inversely proportional the lifetime of the excited state). In order to accurately determine peak profiles, narrow scans were taken near many of the peaks with the pump laser seeded to reduce the OPA bandwidth to ~0.01 meV. Examples of narrow range scans of individual peaks with fits are shown in Figures 4 and 5. The measured resonance energies and widths for Peaks 13 - 23 are listed in Table I (the measured parameters for Peaks 1 - 12 were previously reported in Ref. [1]).  The narrowest peaks (13 and 18) are instrumentally broadened by the linewidth of the OPA (~0.01 meV), so the widths listed for these peaks are upper limits rather than their natural widths.  Finally, note that there might be additional weak, very broad, or very narrow peaks in the spectrum that are not discerned due to signal-to-noise limitations or overlap with stronger peaks; as examples, there may be structure in the low energy tail of Peak 20 and there is a broad hump centered near 726 meV (under Peaks 17 and 18) that might be due to another resonance or it might be caused by a variation in the continuum photodetachment cross section.

The quoted uncertainties in peak energies include statistical uncertainties associated with the fits, photon energy calibration and bandwidth uncertainties, and potential Doppler shifts due to possible deviation of the ion and laser beam intersection angle from perpendicular (estimated to be within ±5° of perpendicular, which gives a Doppler shift uncertainty of 0.02 meV at a photon energy of 600 meV).  The uncertainties in the peak widths only include the fitting uncertainties.

In the present case, all of the peaks except Peak 19 were found to be highly symmetric with large asymmetry parameters ($|q| > \sim 10$), so that the Fano fits were almost indistinguishable from Lorentzian fits of those peaks.  For example, Fig. 4 shows two closely spaced peaks (13 and 14) that display the symmetric profiles. This finding of highly symmetric peaks is to be expected for Feshbach resonances embedded in a continuum in which the resonance contribution to the cross section is much stronger than the continuum component, thus making interference effects minimal.  In contrast, Peak 19 (see Fig. 5) has a characteristic "Fano resonance" profile with a small asymmetry parameter of $q = 0.75(15)$, indicating significant interference between the resonant and continuum photodetachment channels.

Patterns in the separations of the peaks permit identification of the lower states and the energies of the upper states for some of the observed transitions. If two lower states both make transitions to the same upper state, the resulting spectrum shows a pair of peaks that are separated by the known energy interval between the lower states. For example, Peaks 14 and 17 are separated by an energy difference of 83.93(5) meV, which closely matches the previously determined fine structure splitting between the La$^-$ $^3F_3^e$ and $^3F_2^e$ bound states of 83.941(20) meV [1,11,12]. Therefore, the lower states for Peaks 14 and 17 are identified as the $^3F_3^e$ and $^3F_2^e$ states, respectively, and these two peaks result from transitions to the same quasibound upper state, designated as "$A$". Similarly, Peaks 15 and 19 are separated by approximately this same interval, 83.88(12) meV, indicating that they arise from transitions from the same two lower states $^3F_3^e$ and $^3F_2^e$, respectively, but to a different quasibound upper state, $B$. Peaks 13 and 18 are separated by an energy difference of 88.91(4) meV, which closely matches the previously determined fine structure splitting between the $^3F_4^e$ and $^3F_3^e$ bound states of 88.928(33) meV [1,11,12]. Therefore, Peaks 13 and 18 are identified as due to transitions from the $^3F_4^e$ and $^3F_3^e$ states, respectively, to a third quasibound upper state, $C$. Similarly, Peaks 20 and 23 are separated by approximately this same interval, 89.3(23) meV, indicating that they arise from transitions from the same two lower states $^3F_4^e$ and $^3F_3^e$, respectively, but to a fourth quasibound upper state, $D$. These transition assignments are shown in Fig. 1 and listed in Table I. Note that the measured widths of each of the pairs of peaks that reach the same upper state are the same within uncertainties, as expected since the peak width depends on the lifetime of the upper state. Finally, since the lower state energies are known for the transitions reaching the quasibound states $A$, $B$, $C$, and $D$, the energies of these states can be determined from the present measurements; these state energies are given in Table II.

The measured La⁻ quasibound state energies and peak widths give clues to the characters of the observed resonances. The quasibound states *A*, *B*, and *C* lie at excitation energies above the neutral La ground state of 165.79(5) meV, 203.71(7), and 253.72(5) meV, respectively. This places them in energy between the neutral atom ground state La $^2D_{3/2}$ and first excited electronic state La $^4F_{3/2}$ (see Fig. 1). This location of these states, coupled with the narrowness of the peak transitions reaching them (widths < 1 meV), suggests that *A*, *B*, and *C* are Feshbach resonances below the neutral atom first excited electronic state manifold La ($5d^26s$ $^4F_J$). Given that the transitions reaching them all originate from even-parity $5d^26s^2$ lower states, these resonances are most likely odd-parity $5d^26s6p$ quasibound states. Effectively, they are formed by attaching a $6p$ electron to a La ($5d^26s$ $^4F_J$) excited state. After photoexcitation, these resonances subsequently autodetach to leave the neutral atom in the ground electronic state La ($5d6s^2$ $^2D_J$). This would be a two-electron decay process; therefore, the autodetachment lifetime should be relatively long resulting in the narrowness of the peaks. Peak 16 may also be of this same Feshbach character, given its narrow width of 0.042(11) meV, however in the present experiments it is not possible to identify the lower state or determine the energy of the upper state for this resonance.

In contrast to the narrow peaks in the lower energy range, the four higher energy Peaks 20 - 23 are considerably wider (> 5 meV), indicating that these resonances have shorter lifetimes which would be more typical of shape resonances rather than Feshbach resonances. In addition, the energy of the upper state for Peaks 20 and 23, the quasibound state *D*, is 422.7(11) meV, which places it above the $^4F_{5/2}$ excited state of La at 373.1927 meV [22]. Therefore, Peaks 20 - 23 are most likely due to shape resonances positioned above thresholds in the La ($5d^26s$ $^4F_J$) excited state manifold. Similar cases of negative ion shape resonances above neutral atom thresholds have been observed in a number of previous photodetachment experiments, for example in He⁻ [27,28], Ca⁻ [29], and Cs⁻ [30].

Finally, we note that Peaks 15 and 19 show an interesting contrast in their resonance profiles (see Fig. 5). Both of these peaks have the same upper state, $B$, but they have different lower states $^3F_3^e$ and $^3F_2^e$, respectively. Whereas Peak 15 shows a large symmetric increase in the photodetachment cross section, Peak 19 is considerably weaker and shows a characteristic Fano interference profile with a decrease in the cross section followed closely by an increase. The distinctive shape of Peak 19 is caused by destructive and constructive interference between the resonant and continuum detachment channels as the relative phase changes across the range of the resonance. In contrast, Peak 15 only shows a positive contribution to the cross section, because the resonant channel is substantially stronger than the continuum channel in that case thus the interference of the two channels is negligible for this transition.

## **IV. Summary**

In summary, we have used tunable spectroscopy to investigate the photodetachment spectrum from La¯ in the region above the neutral La detachment threshold. The spectrum shows multiple thresholds that are identified as due to the opening of photodetachment channels between different states of La¯ and ground and excited states of La. At least 11 distinct resonance peaks are observed over the photon energy range 590 - 920 meV. Three quasibound negative ion states giving photodetachment resonances of Feshbach character are identified and their energies measured through multiple transitions that excite them from different lower bound lower states of La¯. In addition, four shape resonances are observed above the first excited electronic state manifold of neutral La.

The present investigation extends our group's previous study of the bound state structure of La¯ [1] to provide detailed information about the photodetachment region above the neutral atom threshold. Further insights about the observed resonances will require theoretical investigations of the structures and energies of La¯ quasibound states. These results can be used to test theoretical calculations for this heavy negative ion that include detailed accounting of multielectron correlation and relativistic effects. Such further investigations are needed to fully evaluate La¯ as a potential candidate for laser-cooling applications.

We thank Lin Pan for useful discussions, Marianna Safronova for valuable insights, and Dave Burdick and Sam Strosnider for technical assistance. This material is based in part upon work supported by the National Science Foundation under Grant Nos. 1404109 and 1707743. N. L. and J. W. received partial support from Denison University's Anderson Summer Research Fund.

**TABLE I**. Measured peak energies and widths (with 1 $\sigma$ uncertainties in parentheses) for the 11 La¯ photodetachment resonances observed in the present study; Peaks 1 - 12 were previously characterized by Walter *et al.* [1]. The narrowest peaks (13 and 18) are instrumentally broadened by the linewidth of the OPA (~0.01 meV), so the widths listed for those peaks are upper limits rather than their natural widths. Several of the peaks have transition assignments listed from bound states of La¯ to quasibound excited states, as discussed in the text.

| Peak | Energy (meV) | Width (meV) | Transition |
|------|--------------|-------------|------------|
| 13 | 638.41(3) | 0.022(6) | $^3F_4^e \to C$ |
| 14 | 639.41(5) | 0.37(7) | $^3F_3^e \to A$ |
| 15 | 677.36(5) | 0.85(13) | $^3F_3^e \to B$ |
| 16 | 701.01(4) | 0.042(11) | |
| 17 | 723.34(6) | 0.34(5) | $^3F_2^e \to A$ |
| 18 | 727.32(3) | 0.018(6) | $^3F_3^e \to C$ |
| 19 | 761.24(9) | 0.72(13) | $^3F_2^e \to B$ |
| 20 | 806.3(13) | 11(3) | $^3F_4^e \to D$ |
| 21 | 847.8(9) | 6.2(10) | |
| 22 | 872.1(12) | 7.6(19) | |
| 23 | 895.6(19) | 8.8(18) | $^3F_3^e \to D$ |

TABLE II. Measured excitation energies of quasibound excited states of La¯ relative to the ground state of La ($^2D_{3/2}$) (with 1 $\sigma$ uncertainties in parentheses).

| State | Energy (meV) |
|---|---|
| *A* | 165.79(5) |
| *B* | 203.71(7) |
| *C* | 253.72(5) |
| *D* | 421.7(11) |

**FIGURE CAPTIONS**

**FIG. 1.** Partial energy level diagram showing relevant states of La¯ and La. The excitation energies of La states are from Ref. [22] and the binding energies of bound states of La¯ are from Refs. [1,11,12]. The numbered arrows indicate assigned resonance transitions observed in the present study to the La¯ quasibound states $A$, $B$, $C$, and $D$ (dashed lines).

**FIG. 2.** Measured neutral atom signal for photodetachment from La¯ showing 22 of the 23 observed resonance peaks (Peak 1 is near 260 meV). The data below 585 meV are from Walter *et al.* [1] and the data above 585 meV were acquired in the present study; the vertical scales of the two data sets are adjusted so that the neutral signals match up near their common boundary energy of 585 meV. The data over the entire range has been broadly-binned (~2-5 meV), and the narrow peaks are shown with finer bins. The solid line connects the data points to guide the eye. Relevant photodetachment thresholds from the four most strongly-bound states of La¯ to low-lying states of neutral La are indicated by vertical bars.

**FIG. 3.** Photodetachment spectrum showing the threshold for La¯ ($5d^26s^2\ ^3F_4^e$) —-> La ($5d^26s\ ^4F_{7/2}$) and a nearby resonance Peak 21; the solid line is a fit of a Wigner *p*-wave plus a Fano peak function. The arrow indicates the detachment threshold $E_t$.

**FIG. 4.** Expanded view of two closely spaced resonances, Peaks 13 and 14, with Lorentzian fits.

**FIG. 5.** Measured photodetachment signals from La¯ over a) Peak 15 and b) Peak 19; with fits of the Fano resonance function (Eq. 2). These two peaks originate from transitions that have the same upper state but different lower states; their profiles are radically different due to the relative

strengths of the resonant and continuum photodetachment channels for each transition. Whereas Peak 15 shows only a constructive contribution to the cross section due to the resonance, Peak 19 shows both destructive and constructive interference in the resonance-continuum interaction.

FIGURE 1

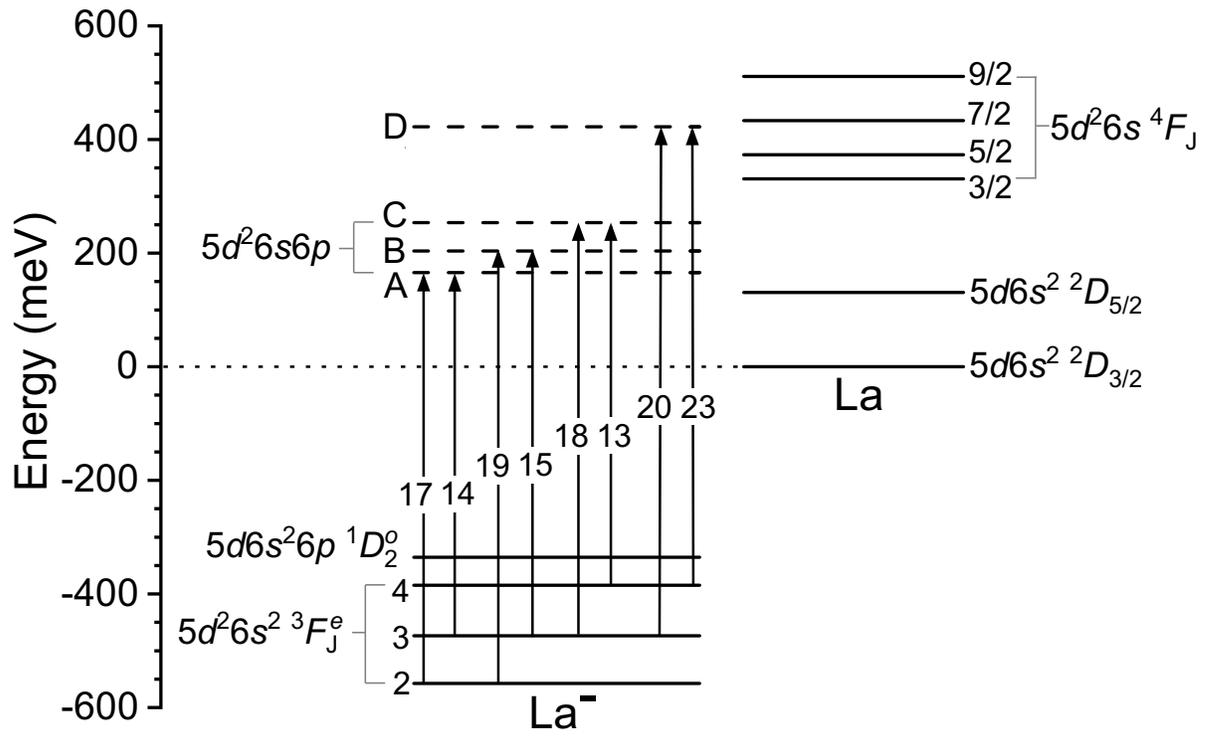

FIGURE 2

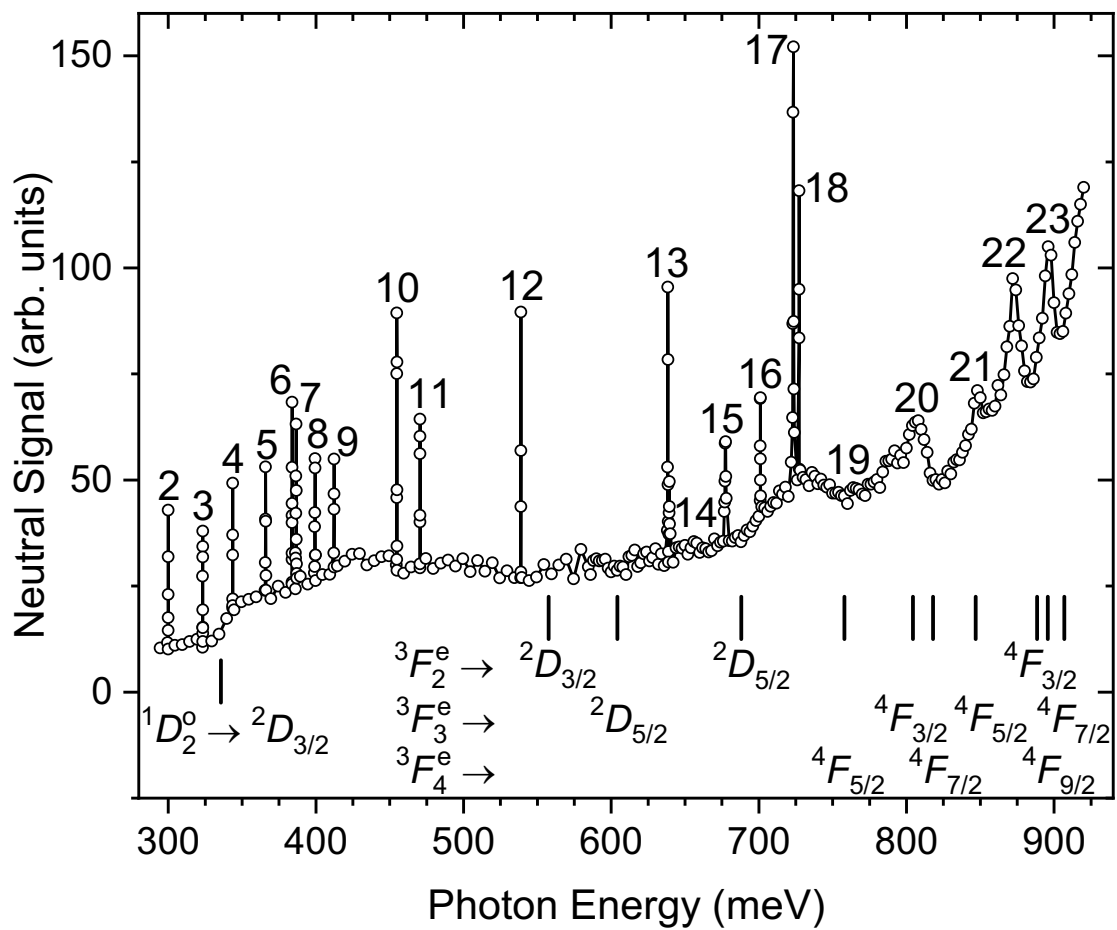

FIGURE 3

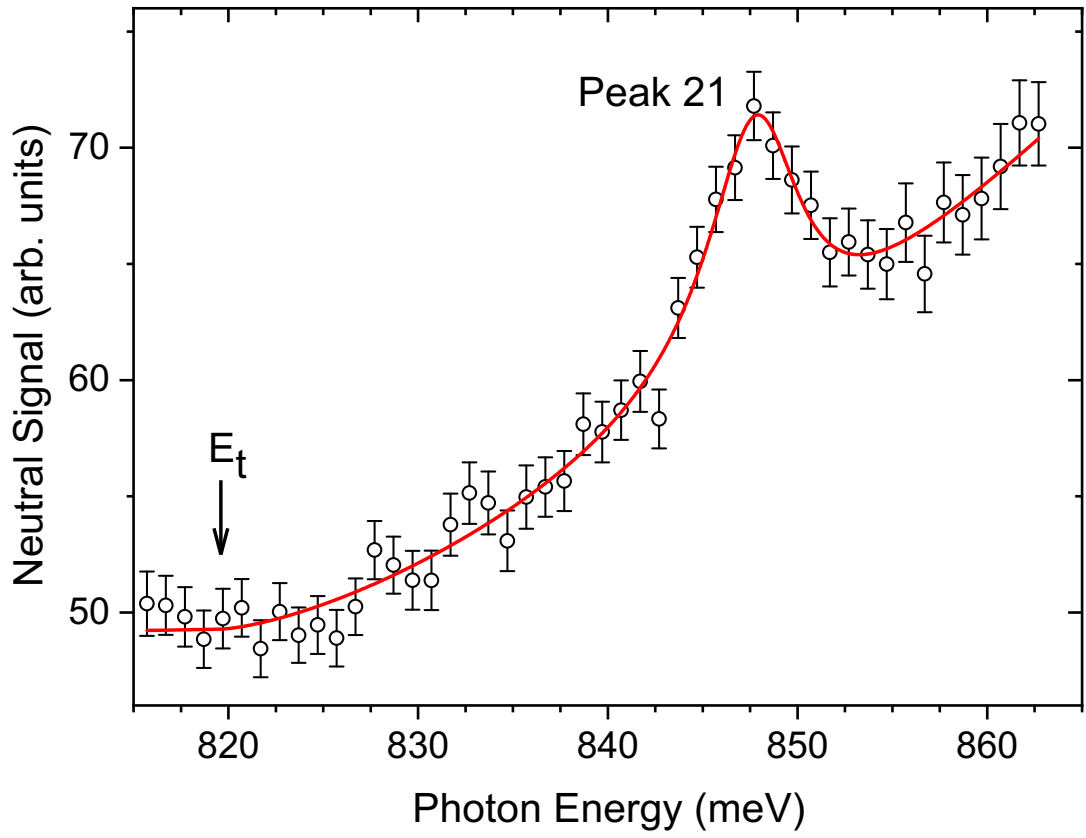

FIGURE 4

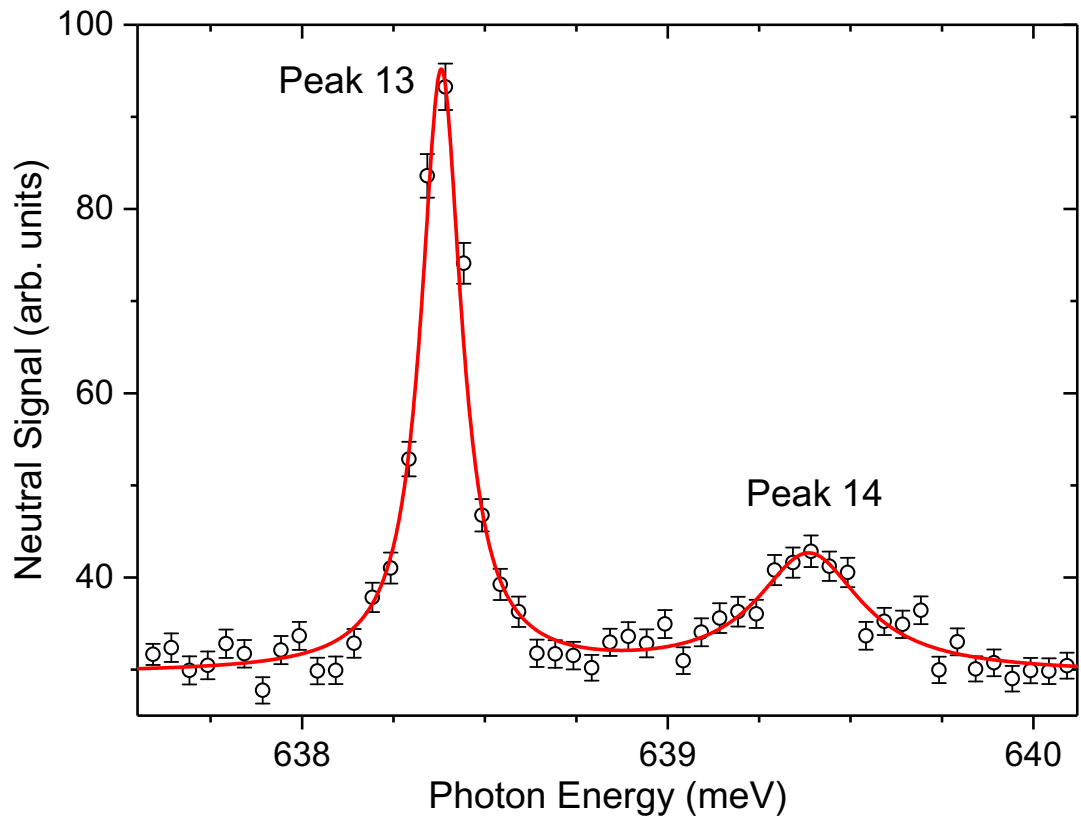

FIGURE 5

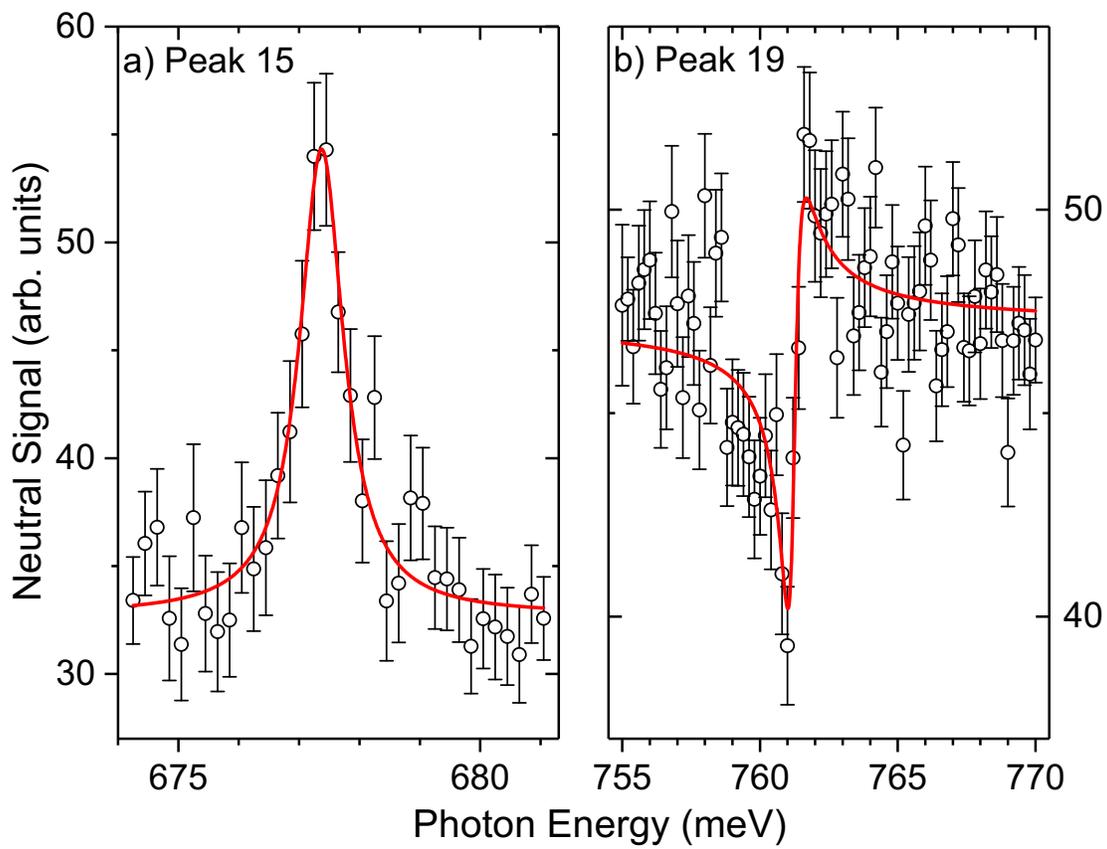